\documentclass[aps,twocolumn,pra,superscriptaddress]{revtex4-2}
\newcounter{defcounter}
\usepackage{graphicx}
\usepackage[space]{grffile}
\usepackage{latexsym}
\usepackage{textcomp}
\usepackage{longtable}
\usepackage{tabulary}
\usepackage{booktabs,array,multirow}
\usepackage{amsfonts,amsmath,amssymb}
\usepackage{natbib}
\usepackage{xr}
\usepackage{siunitx}
\usepackage[version=4]{mhchem}
\newcommand{\be}{\begin{equation}}
\newcommand{\ee}{\end{equation}}
\newcommand{\bi}{\begin{itemize}}
\newcommand{\ei}{\end{itemize}}
\newcommand{\bea}{\begin{eqnarray}}
\newcommand{\eea}{\end{eqnarray}}
\newcommand{\Eq}[1]{Eq.(\ref{#1})}
\newcommand{\Fig}[1]{Fig.\,\ref{#1}}
\newcommand{\Sec}[1]{Sec.\,\ref{#1}}
\newcommand{\Onlinecite}[1]{Ref.\nocite{#1}\citenum{#1}}

\newcommand{\FSC}{FSC$^3$}

\newcommand{\Comp}{\ensuremath{{\cal C}}}

\newcommand{\Es}{\ensuremath{E_{\rm s}}}
\newcommand{\Ec}{\ensuremath{E_{\rm c}}}

\newcommand{\sigt}{\ensuremath{\tilde{\sigma}}}
\newcommand{\sigtt}{\ensuremath{\tilde{\sigma}_\mathrm{t}}}

\newcommand{\class}[1]{\ensuremath{\mathrm{c}_{#1}}}

\newcommand{\Ns}{\ensuremath{N_\mathrm{s}}}

\newcommand{\fm}{\ensuremath{f_\mathrm{m}}}

\newcommand{\fl}{\ensuremath{f_\mathrm{l}}}

\newcommand{\iv}{{\it I-V}}

\newcommand{\PrO}{PrO$_x$}
\newcommand{\RuO}{RuO$_2$}

\begin{document}
	\title{Low-energy electron microscopy intensity-voltage data -- factorization, sparse sampling, and classification} 
	\author{F. Masia}\email{masiaf@cf.ac.uk}\affiliation{School of Biosciences, Cardiff University, Museum Avenue, Cardiff CF10 3AX, United Kingdom}\affiliation{School of Physics and Astronomy, Cardiff University, The Parade, Cardiff CF24 3AA, United Kingdom}
	
	\author{W. Langbein}\affiliation{School of Physics and Astronomy, Cardiff University, The Parade, Cardiff CF24 3AA, United Kingdom}
	
	\author{S. Fischer}\affiliation{Institute of Solid State Physics, University of Bremen, Otto-Hahn-Allee 1, 28359 Bremen, Germany}
	\author{J.-O. Krisponeit}\affiliation{Institute of Solid State Physics, University of Bremen, Otto-Hahn-Allee 1, 28359 Bremen, Germany}\affiliation{MAPEX Center for Materials and Processes, University of Bremen, 28359 Bremen, Germany}
	\author{J. Falta}\affiliation{Institute of Solid State Physics, University of Bremen, Otto-Hahn-Allee 1, 28359 Bremen, Germany}\affiliation{MAPEX Center for Materials and Processes, University of Bremen, 28359 Bremen, Germany}
	\date{}

\begin{abstract}
	Low-energy electron microscopy (LEEM) taken as intensity-voltage (\iv) curves provides hyperspectral images of surfaces, which can be used to identify the surface type, but are difficult to analyze. Here, we demonstrate the use of an algorithm for factorizing the data into spectra and concentrations of characteristic components (\FSC) for identifying distinct physical surface phases. Importantly, \FSC\ is an unsupervised and fast algorithm. As example data we use experiments on the growth of praseodymium oxide or ruthenium oxide on ruthenium single crystal substrates, both featuring a complex distribution of coexisting surface components, varying in both chemical composition and crystallographic structure. With the factorization result a sparse sampling method is demonstrated, reducing the measurement time by 1-2 orders of magnitude, relevant for dynamic surface studies. The \FSC\ concentrations are providing the features for a support vector machine (SVM) based supervised classification of the types. Here, specific surface regions which have been identified structurally, via their diffraction pattern, as well as chemically by complementary spectro-microscopic techniques, are used as training sets.  A reliable classification is demonstrated on both exemplary LEEM \iv\ datasets.
\end{abstract}
\maketitle{}
\section{Introduction}

\begin{figure*}
	\centering
	\includegraphics[width=\textwidth]{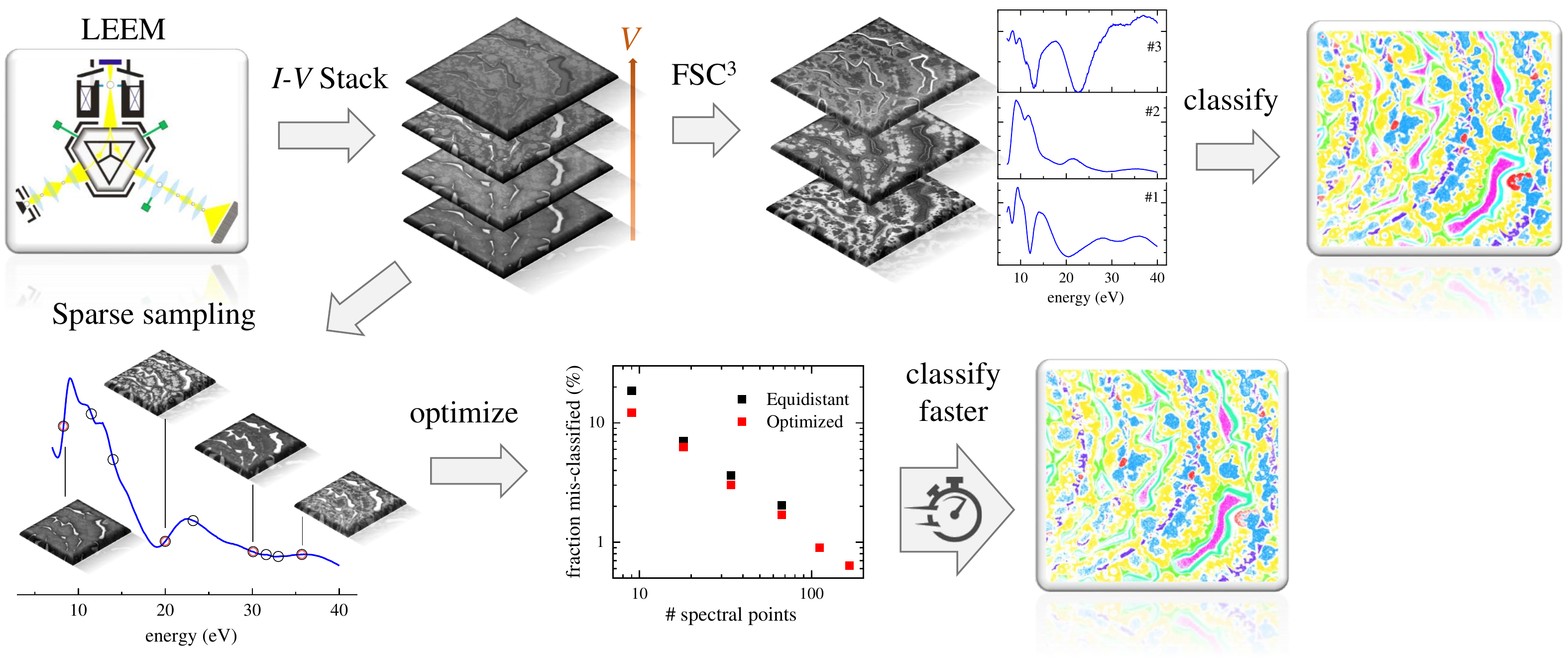}
	\caption{Sketch of the analysis method. LEEM \iv\ stacks are recorded, and factorized into non-negative components and spectra by \FSC. Manually selected regions are used as a training set for a classifier on the resulting component concentrations. Sparse sampling retrieves the \FSC\ concentrations from a smaller number of spectral points, down to the number of \FSC\ components. The spectral positions are optimized for the smallest number of unclassified or misclassified points, resulting  in a measurement speed-up by an order of magnitude.}
	\label{fig:Sketch}
\end{figure*}

Low-energy electron microscopy (LEEM) is a powerful experimental method that provides high-resolution hyperspectral data when used in the \iv\ imaging mode. In many applications, the information sought is the spatial distribution of surface types, i.e. the structure of the first few atomic layers of the investigated sample surface. This, as well as the electronic band structure, is encoded in the energy dependent electron reflectivity of the surface, so called intensity versus electron energy, or in short LEEM  \iv\ curve. Differing from conventional low-energy electron diffraction intensity-voltage analysis (LEED  \iv), where a number of diffracted beams are recorded and analyzed, LEEM  \iv\ analysis is usually restricted to the specular beam, i.e. the (00) reflection in LEED, since excitation of higher order LEED reflections is not possible at low electron energies due to the small radius of the Ewald sphere given by the electron wavevector.

Importantly, the resulting LEEM  \iv\ spectra are difficult and time-consuming to interpret. The dominating diffraction contrast is well understood for higher energies in the context of conventional LEED  \iv\ analysis.\cite{pendry1974,hove1979} However, the electron energies accessible in LEEM are in the so-called VLEED (very low energy electron diffraction) region at energies below \SI{50}{\eV}. For these energies, the interaction of the electrons with the surface cannot be understood in terms of the dynamic diffraction theory which is successfully applied for higher energies.\cite{pendry1974,hove1979}  Instead, the theoretical description is complex and electron band interactions must be taken into account.\cite{pendry1969,krasovskii2004,flege2014,krasovskii2014} Moreover, for a complete description of the data, additional contrast mechanisms like phase contrast at steps need to be considered \cite{altman1998,chung1998,pang2009} and calculations for surfaces with large and inhomogeneous surface structures are not available. To circumvent this problem, a fingerprinting approach has been established already in the early 1990s.\cite{over1993,over1994,schmid1995} The term fingerprinting refers to comparing measured data to well known spectra (\iv\ curves). The knowledge on such spectra being a characteristic signature of a specific surface phase is usually established by employing complementary methods like X-ray photoemission electron microscopy (XPEEM), providing chemical information, or LEED reference data of homogeneous surface regions. A typical task in an LEEM \iv\ experiment is however the study of unknown surface phases which need to be identified without such prior knowledge. This is the starting point of the present study.

The task at hand is sorting the measured spectra into groups based on their similarity, at best with pixel resolution. When monitoring dynamic changes of the surface, e.g. by growth processes or during surface reactions, this classification should be fast and allow for real-time data acquisition and analysis. Capturing a hyperspectral image stack with high energy resolution and sufficient signal to noise typically takes several \SI{10}{\minute}, which is often too slow to follow the growth dynamics. Instead, a faster operation is required, which with present instrumentation can only be achieved by reducing the spectral density of the measurement, i.e. reducing the number of energies measured.

In the literature, classification has been done using principal component analysis of the \iv\ spectra, to reduce the dimensionality of the single pixel data, followed by unsupervised clustering algorithms like $k$-means \cite{dejong2020}. This type of classification algorithm a priori assumes that each pixel is representing only one phase. Mixed phases, however, which are collected as a superposition of their components into one pixel spectrum, must be accounted for afterwards either by assignment to the most similar ``majority'' phase, or by leaving such pixels unclassified.

We show here a method (see sketch in \Fig{fig:Sketch}) which factorizes the LEEM hyperspectral data into non-negative components and concentrations, using the \FSC\ algorithm.\cite{MasiaAC13,MasiaJRS15,KarunaJRS16} Using the concentrations, a supervised classification is developed using pair-wise support vector machines (SVM) taking into account the uncertainty of the training set to evaluate a classification probability. This method provides a fast and reliable classification of surface reconstructions, as we show in two examples, ruthenium oxide (\RuO), and praseodymium oxide (\PrO). 

Furthermore, using the extracted component spectra, we demonstrate a sparse sampling (SS) method similar to our previous work,\cite{MasiaOE14} but using the \FSC\ component spectra for projection and the resulting concentrations for classification. We show that even reducing the number of spectral points to the number of components, which in our example of PrOx  are 9 components, down from 235 in the original spectra, more than 90\% of points were classified, of which less than 10\% were misclassified compared to the result using all spectral points, while reducing the acquisition time by a factor of 30 per classification for the present conditions. We note that the performance is dominated by the reduction in signal to noise ratio (SNR) due to the reduced number of spectral points. The SNR is of the order of 20 for each point, limited by the shot noise of the of the order of 1000 electrons detected per pixel, and the typical noise factor of the MCP amplifier of about two. \cite{MoranSPIE97} We therefore expect that taking data with higher SNR, for example by a larger integration time or a higher beam current would improve the results of the SS classifier, even when using the minimum number of spectral points.


This article is structured as follows: Section \ref{sec:method} provides a short explanation of the LEEM \iv\ technique, the physical interpretation of \iv\ spectra and their use in fingerprinting approaches. In section \ref{sec:materials}, the two exemplary datasets analysed in this study are introduced and the respective experimental details are provided. In section \ref{sec:analysis}, the data analysis procedure is presented in detail, including the pre-processing of the data, the factorization approach, the classification technique and the optimization by sparse sampling. The results obtained on the exemplary datasets are presented in section \ref{sec:results}, followed by the conclusions.

\section{LEEM \iv\ Methodology}
\label{sec:method}
Low-energy electron microscopy (LEEM) is a versatile technique used in surface science. It provides means for \emph{in situ} structural and electronic characterization of surfaces in ultra-high vacuum. An electron beam with an electron energy of typically a few eV to a few tens of eV is used to homogeneously illuminate the imaged region of the sample surface.
The elastically backscattered electrons pass a lens system and are projected onto the detector, forming an image. This full-field microscopy technique enables a high time resolution of the order of 100\,ms at a lateral resolution better than \SI{10}{\nano\meter}. Hence, LEEM allows for observing nanoscale surface processes in real time. This can be exploited, for instance, to study surface reactions in gas atmospheres up to 1$\times$10$^{-4}$ Torr as well as growth mechanisms in molecular beam epitaxy (MBE).\cite{bauer1994,flege2012}

A key feature of LEEM is that it can be operated in a spectro-microscopic mode: When the electron energy is varied in small steps across a given range, a distinct intensity-voltage spectrum (\iv\ curve) can be acquired for each individual pixel of the detector camera. The \iv\ spectrum provides the energy-dependent reflectivity of the sample surface, where contrast mainly arises from the atomic structure and the electronic band structure. At step edges, phase contrast can additionally occur, and larger objects can cause image distortions. Depending on the diffracted order that is used for imaging, it is called bright field (using the specular (00) reflection), or dark field imaging (using other reflections).

\subsection{Physical interpretation of \iv\ spectra}

In general, solving the Schrödinger equation for the imaging electrons over the whole space of vacuum and crystal is required to generate theoretical \iv\ curves. This is needed in order to account for effects of both diffraction and electronic band structure on the electron reflection. State-of-the-art \emph{ab initio} calculations, using a Bloch wave ansatz in the crystal half space, are able to reproduce experimental spectra.\cite{flege2014,krasovskii2014} However, since these calculations are complex and require an accurate crystallographic model of an assumed phase \emph{a priori}, the backward problem, i.e. the structural and electronic characterization of unknown phases solely from their specific experimental \iv\ spectra, must be considered highly complex. As mentioned before, to bypass this, \iv\ curves can be used as a fingerprint for distinct surface phases. \cite{flege2018} In this way, established fingerprints can be used to identify the phase of regions as small as the spatial resolution of the imaging.

\subsection{Fingerprinting and classification approaches}

LEEM offers several methods which allow for an assignment of \iv\ curves to structure: By adjusting the electron optics between sample and detector, it is possible to project the LEED pattern from the back focal plane of the objective lens onto the detector. Using an aperture, the surface area that constitutes the LEED pattern can be restricted down to diameters of sub-micrometer dimensions. Thus, if large enough regions of the surface phase in question exist, information about the crystal structure can be inferred and taken into consideration for determining the phase. Also, a LEEM instrument can used in emission mode by exciting the surface with light, typically using a synchrotron as photon source to achieve tunable high intensity illumination, making  spectro-microscopic methods available via photoelectron emission (x-ray photoemission electron microscopy, XPEEM) \cite{schmidt1998}. The analysis of local x-ray photoelectron and absorption spectra allows for a chemical analysis of the studied surface phase.

This assignment allows to establish \iv\ fingerprints for a range of surface phases. However, the identification of surface regions based on these fingerprints is often far from trivial. Firstly, the image stacks depicting a fine-grained composition of species contain a considerable amount of edge pixels, where spectra of neighboring phases are superimposed. Secondly, typical experiments may involve a coexistence of numerous surface phases distributed with very different area fractions. Frequently, new phases, having unknown \iv\ spectra, occur in addition to the established ones. Hence, the first challenge usually consist of dissecting the data stack into an unknown number of components, as many as there are physically distinct phases present at the surface, and elaborate their characteristic spectra therefrom.

\defcitealias{dejong2020}{De Jong et~al.}
\citetalias{dejong2020}\cite{dejong2020} presented a method where the spectra are first reduced in dimension by principal component analysis, after which a $k$-means clustering algorithm is employed to classify each pixel. The spectra compiled by this can then be compared to the known fingerprints to verify them. Significantly though, it is not trivial to separate an unknown number and distribution of phases in the spectral space. Because no prior information is taken into account, edge cases are not handled properly when phases superpose or when two phases have a smaller spectral variation from one another than what a third phase might encompass in itself. Also, because the $k$-means algorithm does not employ any statistical model, evaluating the confidence in the classification requires additional analysis. 

\section{Experimental data}\label{sec:materials}

Here we provide details of the experimental data sets used in the analysis as exemplary data.

\subsection{Praseodymium Oxide}

The \PrO\ LEEM \iv\ data was recorded on an ultrathin praseo\-dymium oxide film on a Ru(0001) surface, using the Elmitec SPELEEM at the I311 beamline\cite{zakharov2012,nyholm2001} of the MAX-lab synchrotron radiation facility in Lund, Sweden, as described in detail in \Onlinecite{flege2017,hocker2017}. It shows a complex band-like morphology of coalesced oxide islands, arranged along the step edges of the substrate.   The Ru substrate, which is covered with an oxygen adlayer, can be identified reliably based on its characteristic \iv\ fingerprint. The praseodymia bands, on the other hand, comprise a rich substructure of coexisting  surface species. Structural information was obtained from the individual phases via \textmu-LEED, and x-ray absorption spectra of the same region were collected in PEEM mode. Finally, in good agreement with theoretical reflectivity curves, five distinct praseo\-dymium oxide phases have been characterized, with differences in stoichiometry, crystallographic structure and even surface termination. While this dataset here serves to illustrates the complexity that can arise in such growth experiments, and presents a challenging test case for the presented numerical approach, the reader is referred to the original studies \cite{flege2017,hocker2017} for a detailed physical description of this surface system. 

\subsection{Ruthenium Oxide}

The \RuO\ data set was recorded using the Elmitec SPELEEM at the Institute of Solid State Physics at the University of Bremen. The instrument is equipped with a detection assembly of multi-channel plates and a phosphorous screen by PHOTONIS and a pco.1600 cooled CCD camera by PCO for data acquisition with low readout noise. A set of  illumination apertures allows micro-diffraction measurements on areas as small as \SI{250}{\nano\meter} in diameter.	

The sample was prepared \emph{in situ} in the LEEM instrument. First, the substrate, a commercial \ce{Ru(0001)} single crystal by Mateck (\ang{<0.1} miscut), was cleaned by repeatedly oxidizing the surface with molecular oxygen and then flash-annealing it to \SI{>1400}{\celsius}. \cite{madey1975} 
It was then subjected to atomic oxygen from an OBS 40 thermal cracker (Dr. Eberl MBE-Komponenten GmbH). The cracker was operated at \SI{1780}{\celsius}; given the chamber geometry and the manufacturer's information, a cracking efficiency of \SI{15}{\percent} is assumed. The total atomic oxygen dose used over a duration of ca. \SI{180}{\minute} thus amounts to \SI{1250}{L}. The sample temperature was kept at \SI{415}{\celsius} during the oxidation process. After this preparation, the \iv\ LEEM image stack was acquired. For each energy in the range of \SI{3}{\eV}=\SI{50}{\eV} with a step size of \SI{0.2}{\eV}, a bright field image was taken with an exposure time of \SI{4}{\second}.

\section{Data analysis}\label{sec:analysis}

LEEM \iv\ image stacks were processed using the analysis pipeline detailed in the following.

\subsection{Image pre-processing}

At very low kinetic energies, the electron trajectories are strongly deflected at topographical features. In the analysis we hence disregard images recorded in this energy regime.

For the \PrO\ data, containing 331 spectral points, we start by registering the hyperspectral data in-plane to compensate for lateral instrumental drift, using a translation vector between two consecutive spectral frames determined by the Matlab function \textit{imregtform}.
We then denoise the data applying a singular value decomposition (SVD)\,\cite{MasiaAC13} on whitened data (the intensity noise in the data is dominated by electron shot noise, scaling proportional to the square root of the intensity -- we thus use the square root of the data which has a noise independent of intensity, {\it i.e.} it is whitened), and retain only the 50 SVD components of highest value in the reconstructed data.
We then correct for vignetting by fitting the data at 7.8\,eV, which provides the most homogeneous spatial pattern (excluding outliers below 67\% or above 144\% of the mean) with a two-dimensional Gaussian function. The fit is then normalized to unity centre value and the data at all energies are divided by it.

For the \ce{RuO2} data, containing 236 spectral points, we had access to measurements of the system sensitivity and dark current, which we used to correct the inhomogeneous illumination. The sensitivity correction was applied following the registration discussed above. Regions having below 2\% of the maximum sensitivity were removed. After this, SVD denoising is done as above, followed by vignetting correction using the spectrally averaged data above 15\,eV for the fit, excluding pixels with values above $135\%$ of the mean.

\subsection{Factorization}
\label{sec:FSC}
After pre-processing, the hyperspectral data are decomposed as a linear combination of components using \FSC, which is employing a non-negative matrix factorisation (NMF) algorithm. For the \PrO\ dataset the standard algorithm was used, while for the \RuO\ dataset, we disregarded energies below 15\,eV (which were dominated by the substrate surface areas) and used a weighted algorithm iteratively taking into account the spectral \Es\ error to improve retrieval of rare components, as described in \Onlinecite{MasiaJRS15}. Specifically, the weight at iteration step $i+1$ at pixel $k$ is given by 
$$w_{i+1}(k)=\left(w_i(k)\frac{\Es(k)}{\langle\Es\rangle}\right)^{1-\alpha},$$
where the $\langle.\rangle$ denotes the average over the spatial pixels, and we used $\alpha=0.5$. Positions with a resulting weight exceeding $\sqrt[4]{P}$, where $P$ is the total number of pixels in the image, were removed from the analysis in the subsequent iteration. For the \PrO\ (\RuO) dataset, nine (six) components were found to represent the data well, as judged using the spectral error showing a spatial average of around 1\%.

\subsection{Classification}
\label{sec:classi}
The concentrations of the components obtained from the factorisation are used as features for a supervised classification using a Support Vector Machine (SVM). The component spatial distributions are used to identify the areas of a specific class, which are then used as training set for the SVM. For the \PrO\ (\RuO) data, $N=7$ ($N=4$) areas were selected, each defining a class. For each pair of classes $(i,j)$, we calculate a binary SVM classifier using the two corresponding training sets, and use it to determine, for each pixel $k$ in the image, the distance from the corresponding SVM hyperplane, $d_{ij}(k)$. For each pair, we choose a distance scale to provide a mean of +1(-1) for the training sets of class $i(j)$, respectively, and determine the resulting standard deviation $\sigma_{ij}$ of $d_{ij}$ for class $i$.  Assuming a Gaussian distribution, we then calculate a probability density that a pixel $k$ is associated to class $i$ as 
\be \label{eq:pij} p_{ij}(k)=\frac{1}{\sigma_{i,j}\sqrt{2\pi}}\exp\left(-\frac{(d_{ij}(k)-1)^2}{2\sigma_{ij}^2}\right)\,. \ee
The product of this density over all pairs of classes $(i,j)$,
\be P_i(k)=\prod_{j\neq i} p_{ij}(k)\,,\ee
defines a probability density that the pixel is associated to the class $i$.  The class $l$ with the largest probability density, i.e. $P_l(k)\geq P_i(k)$ for all classes $i$, is assigned to the pixel $k$. To quantify the likelihood of the classification, we normalize $P_i(k)$ by the product of the maxima of the probability densities \Eq{eq:pij}, yielding 
\be \tilde{P}_i(k)=\prod_{j\neq i} \sqrt{2\pi}\sigma_{ij} p_{ij}(k) \ee
and define the normalized average standard deviation
\be \sigt_i= \sqrt{\frac{2}{1-N}\log\left(\tilde{P}_i\right)}\ee
providing the average distance of the point from the training set mean in units of the training set standard deviations, quantifying the confidence of the classification. 
We introduce a threshold $\sigtt$, to define points with $\sigt_i>\sigtt$ as unclassified.

\subsection{Sparse sampling}
\label{sec:sparse}
In \Onlinecite{MasiaOE14} we have introduced a method to increase the acquisition speed in sequential hyperspectral imaging based on the concept of sparse sampling. In that work, we have used SVD to define a basis given by the highest singular values, and find the spectral positions of a small number of spectral points minimising the error in the reconstruction using this basis. SVD was used as opposed to NMF to determine the basis as the reconstructed quantity was subject to further non-linear processing prior to representing physically constrained concentrations. Here, we present a corresponding method for the LEEM data.

We use an image which covers the full spectral range sampled at the Nyquist limit of the instrument spectral resolution to determine a subset of the spectral points by minimising a figure of merit (FOM) related to the classification results. The \FSC\ spectra $S$ determined as detailed in \Sec{sec:FSC} are used as a basis for reconstruction. A hyperspectral image $D^*$, acquired at a limited number of \Ns\ spectral points, can be projected into the set of spectra obtained from \FSC\ of the data with full spectral information, thus calculating the concentration distributions $C^*$ which can be used as features for the classifications. The concentrations are determined by solving the system $S^*C^*=D^*$ using a non-negativity constraint, where $S^*$ is given by $S$ taken at the spectral points of $D^*$.
The classification obtained using the data with sparse sampled spectral points can differ from the classification resulting from the analysis of the data with full spectral information. To inform the choice of the spectral points, we define two FOMs; (i) the fraction of points $f_{\rm l}$ with $\sigt>\sigtt$, representing the loss of information; (ii) the fraction of misclassified points $f_{\rm m}$ relative to the classification using the full spectral information. To test the algorithm we have divided the images into a $4\times4$ checker board pattern, where the `black` fields are used to define factorization, classification, and optimization of spectral points, while the `white` fields serve to verify the method on an unseen data set.

\FSC\ factorisation and classification using the full spectral information of the black fields define our ground truth. We then apply a random walk algorithm to optimise the set of $N_{\rm s}$ spectral points as follows. Starting with  equidistant points covering the measured range, each iteration moves the points to a random position within the interval covering half the distance to the adjacent spectral points, in this way conserving the spectral ordering while exploring the whole spectral range. Using the data $D^*$ at the sparse spectral points $S^*$ , we determine the concentration maps $C^*$ as discussed, and classify the points using the classifier obtained from the full spectral information. The FOM's for this classification are evaluated and the new spectral points are kept for the next iteration only if the FOM was reduced. The iteration is stopped after a given number of loops -- chosen to be 500 unless otherwise noted.

For validation, we sample the white fields at the spectral points which minimise the FOM, project and classify using the same classifier, and evaluate its FOM.

\begin{figure*}
	\centering
	\includegraphics[width=\textwidth]{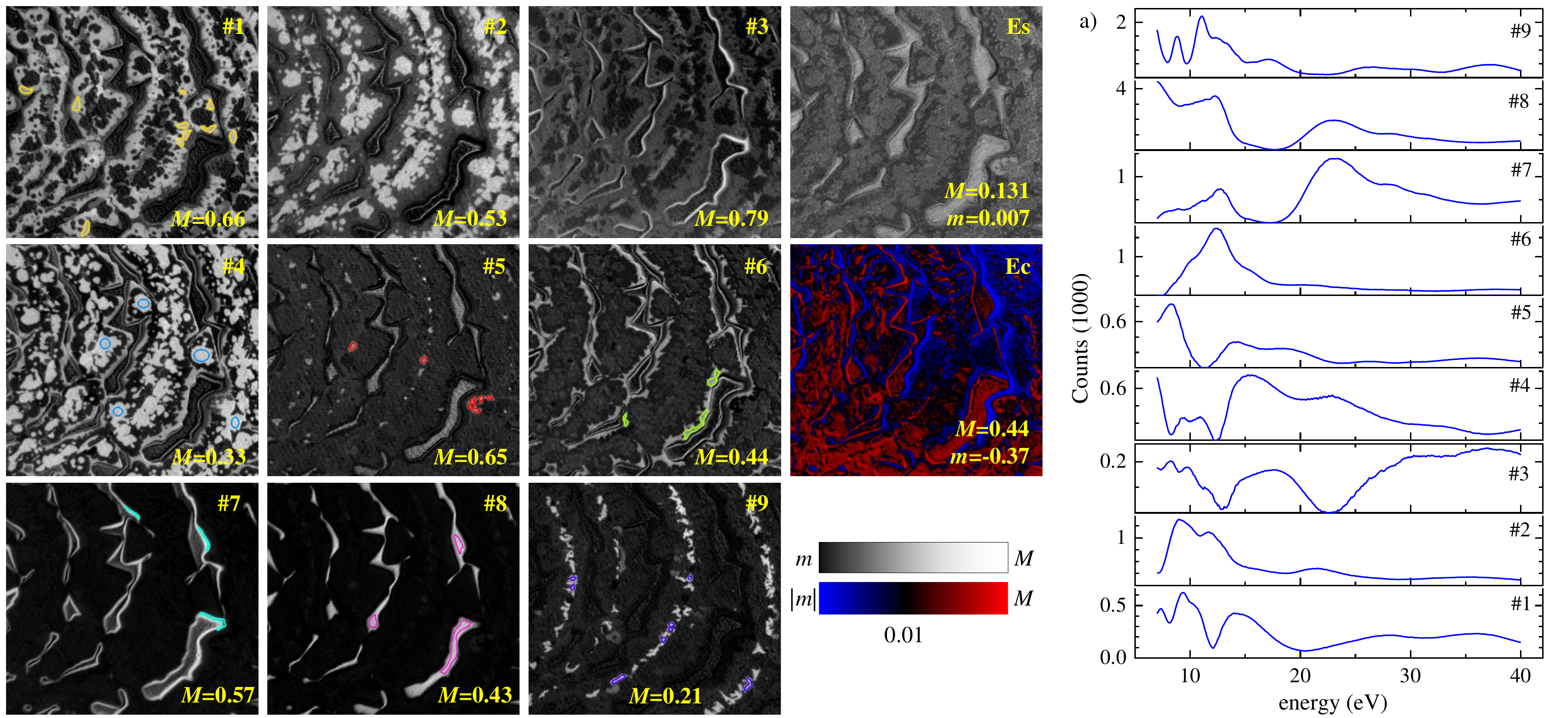}
	\caption{\FSC\ analysis of LEEM \iv\ data taken on \PrO. Concentrations of the components $\Comp_i$, $i=1..9$ are given in the images on a greyscale from $m=0$ to $M$ as labelled, together with the spectral error \Es. For the latter a logarithmic scale has been used between the given $m$ and $M$. The concentration error \Ec\ is shown on a colour scale of red (blue) hue for $\Ec>0$ ($\Ec<0$) as shown. The value is proportional to $\log\left(\left|\Ec\right|\right)$, with black indicating $\left|\Ec\right|\le 0.01$. Each spatial region used as training set for classification is shown in one of the images encircled by coloured lines. a) Component spectra.}
	\label{fig:PrOxFSC3}
\end{figure*}

\section{Results}\label{sec:results}

In this section we present the results of the data analysis. We use the \PrO\ data to demonstrate the method in detail including sparse sampling, and then provide some results for \RuO\ to exemplify the generality of the method. 

\subsection{\FSC\ of \PrO}
\label{sec:fscpro}
\Fig{fig:PrOxFSC3} shows the \FSC\ analysis of the LEEM \iv\ stack on \ce{PrO_x/Ru(0001)} using 9 components. The sample morphology is clearly visible in these components with little noise. As characterised before,\cite{hocker2017} the surface consists of a flat substrate with bands of coalesced oxide islands which nucleated at the atomic step edges of the \ce{Ru(0001)} substrate. In between, the bare substrate is visible in component $\Comp_8$ and a transition region in the vicinity of the island edges is highlighted in $\Comp_3$ and $\Comp_7$. The $\ce{PrO_x}$ regions comprise a complex substructure of five distinguishable phases. In the central region, located directly at the step edges, one phase is formed as rather small, circular cores (prominent in $\Comp_9$). Approaching the island edges, this central component is surrounded by a series of other distinct oxide phases: the innermost surrounding phase is visible in $\Comp_2$ and $\Comp_4$, a further outward one is discernible in $\Comp_1$ and, finally, $\Comp_6$ is located at the island rims. Additionally, a rather sparse phase is identified in $\Comp_5$. This component is also located primarily at the atomic step edges of the substrate as well as in a more extended, crescent-shaped region. 

In a previous analysis,\cite{flege2017} the same dataset was analyzed by comparing the recorded spectra with curves calculated by \emph{ab initio} scattering theory for a set of candidate phases. The correlation between the calculated and observed curves enabled an identification of the relevant surface phases which are visualised in the detected components. These results can be summarized as follows. The cores as well as the surrounding components, here observable in $\Comp_2$, $\Comp_4$, and $\Comp_6$, showed only subtle differences between their respective \iv\ spectra and a reasonable agreement with the theoretical curve of hexagonal \ce{Pr2O3(0001)}. The variations of this oxide phase originate primarily from the existence of two types of terraces which are separated by atomic steps of half a unit-cell height and feature distinct oxygen terminations. The additional separation between the cores and the outer phase prominent in $\Comp_1$ was attributed to a different thickness and possibly also concomitant variations in strain relaxation and distinct surface reconstructions. Located only along the island rims, a highly-oxidized fluorite \ce{PrO2(111)} phase was identified by an \iv\ curve matching well to the expected theoretical spectrum. The additional sparse phase identified in $\Comp_5$ was ascribed to another \ce{Pr2O3} polymorph, the cubic bixbyite-like phase \ce{Pr2O3(111)}, and the spectra of the ruthenium substrate regions agree with a termination by a $(1\times1)$ reconstructed oxygen adlayer. 

The distribution of \FSC\ components already reflects the arrangement of these previously identified phases remarkably well -- in particular considering that no laboriously precalculated spectra were necessary for this analysis.
The spectral error \Es\ is larger in the substrate regions, whereas the concentration error \Ec\ is large at the transition region, mostly on the left-hand side island rims. This asymmetry might be due to an imaging error caused by not positioning the contrast aperture centered in fourier space and does not indicate the existence of an additional physical surface component. In addition, the height change from the substrate to the islands, which was determined to be about $3-4\,\mathrm{nm}$ by AFM,\cite{hocker2017} causes energy-dependent deflections of the electrons. Generally, the concentration error is larger close to phase boundaries, which could be related to the coherent interference between multiple phases not accounted for by a linear mixing of intensities.

\begin{figure*}
	\centering
	\includegraphics[width=\textwidth]{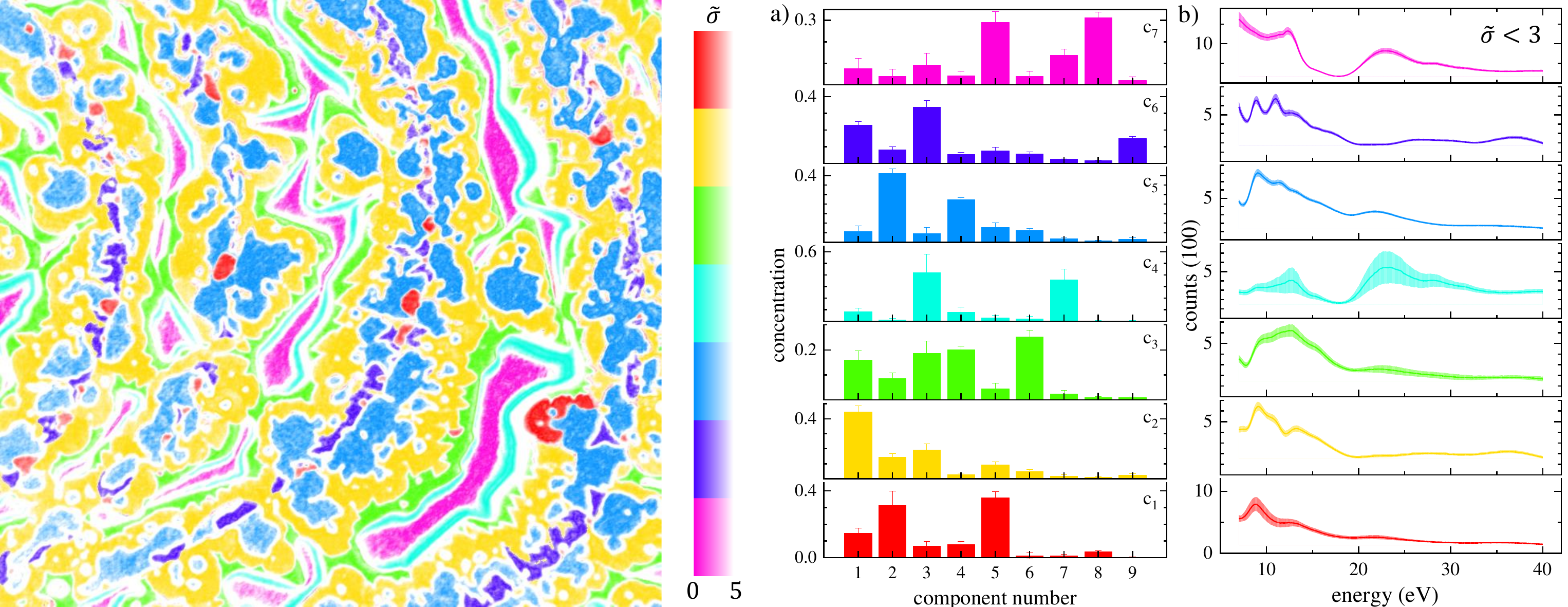}
	\caption{Classification of LEEM \iv\ data taken on \PrO. The hue of the colour represents the assigned class $i$, with a saturation given by $\max\left(1-\sigt_i/5,0\right)$, indicating the assignment confidence. a) Component concentrations with standard deviations taken from the training regions indicated in \Fig{fig:PrOxFSC3} with the same colour coding as the classification results. b) Class LEEM \iv\ spectra (solid lines) considering only the pixels with $\sigt\le3$, with the range of spectra classified into the class given as shaded region. The separate $\sigt$ for each class are given in the SI \Fig{S-fig:PrOxClassSI}.}
	\label{fig:PrOxClass}
\end{figure*}

\begin{figure*}
	\centering
	\includegraphics[width=\textwidth]{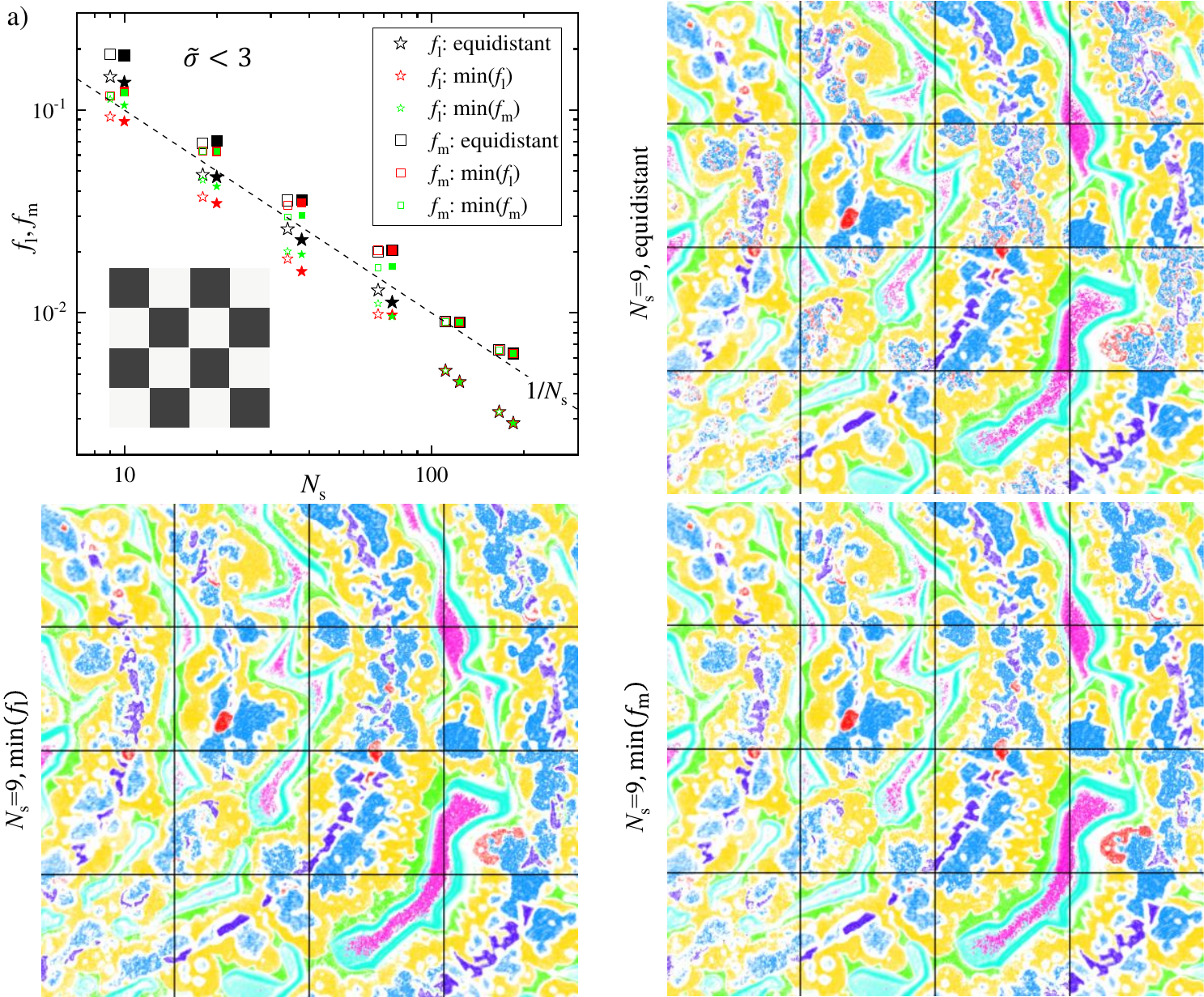}
	\caption{Results of the classification using sparse sampling. a) FOMs \fl\ (stars) and \fm\ (squares) versus the number of spectral points \Ns\ for spectrally equidistant points (black), minimisation of \fl\ (red), or minimisation of \fm\ (green). The checker board black fields are used to define factorization, classification, and optimization of spectral points, while white fields serve to verify the method on an unseen data set. The empty symbols show the FOMs during the optimisation procedure (black fields), while the full symbols (horizontally offset) refer to the FOMs obtained when applying the method in the white validation fields. The latter are slightly displaced on the horizontal axis for visibility. The dashed line shows $1/\Ns$.  The probability threshold used was $\sigtt=3$. The classification images combine the results obtained using the full spectral information (in the black fields) with the results of the sparse sampling (in the white fields) obtained either using \Ns=9 equidistant points or one of the optimisation method. Colour code as \Fig{fig:PrOxClass}. The results for \sigtt\ of 2 and 4 are given in \Fig{S-fig:SparseClassSI}.}
	\label{fig:SparseClass}
\end{figure*}

\subsection{Classification of \PrO}

The \FSC\ results shown in \Fig{fig:PrOxFSC3} are used for a classification as described in \Sec{sec:classi}. The training sets used for the classes are the areas enclosed in the colour-coded outlines shown in selected concentration images in  \Fig{fig:PrOxFSC3}. The resulting classification is given in \Fig{fig:PrOxClass} with classes  \class{i}\ represented by distinct hues, with a saturation encoding the classification confidence \sigt, where $\sigt=0$ (highest confidence) is fully saturated, dropping to zero saturation (white) for $\sigt\ge5$. Again, the spatial structure of the sample is clearly visible, with the different phases discussed in \Sec{sec:fscpro} evident. Notably, the classes are separated by whitish regions of low classification confidence. Classes 7 and 4 are the Ru(0001)-$(1\times1)$-O substrate and the transition regions at the island sidewalls, respectively. There is a rather sharp transition between class 2 and class 3, where the \PrO stoichiometry was found to change,\cite{flege2017} favouring the fully-oxidized fluorite \ce{PrO2} phase at the island rims over the \ce{Pr2O3} of the central regions. Similarly, a sharp transition is observed between class 2 and class 5 on the oxide islands, reflecting the assumed change in the surface termination of the \ce{Pr2O3(0001)}variations. This sharp spatial separation is remarkable as there are only rather subtle differences in the respective spectra. The island centers are consisting of class 1 and class 6, spaced by regions of low classification confidence. With respect to the \FSC\ results in \Sec{sec:fscpro}, an even clearer distinction between the physical surface components could be achieved. This is illustrated by the clear separation of class 3 and class 5, which were both prominent in $\Comp_4$  before.

The component concentration vectors calculated as average and standard deviation over the training regions of \Fig{fig:PrOxFSC3}a are given in \Fig{fig:PrOxClass}a and the resulting average class spectra calculated considering only pixels with $\sigt\le3$ are given in \Fig{fig:PrOxClass}b as lines, together with their range as shaded bands. They show a large variability of \class{4}, the transition region, and \class{1}, the sparse \ce{Pr2O3(111)} phase at the step edges, while the other classes are well defined, specifically \class{2},  \class{5}, \class{6}, and the undisturbed substrate regions \class {7}. The large variability in the transition regions between the substrate and the oxide islands can be attributed to the aforementioned imaging artifacts whose effect is dependent on the electron energy and is gradually decreasing with increasing distance to the edge of the islands, leading to the inhomogeneous transition region. On the other hand, the somewhat higher variability of \class{1}, the bixbyite-like phase, might be ascribed to an incoherent strain state within these regions, which may arise due to the underlying step edges on the substrate.

\begin{figure}
	\centering
	\includegraphics[width=0.5\textwidth]{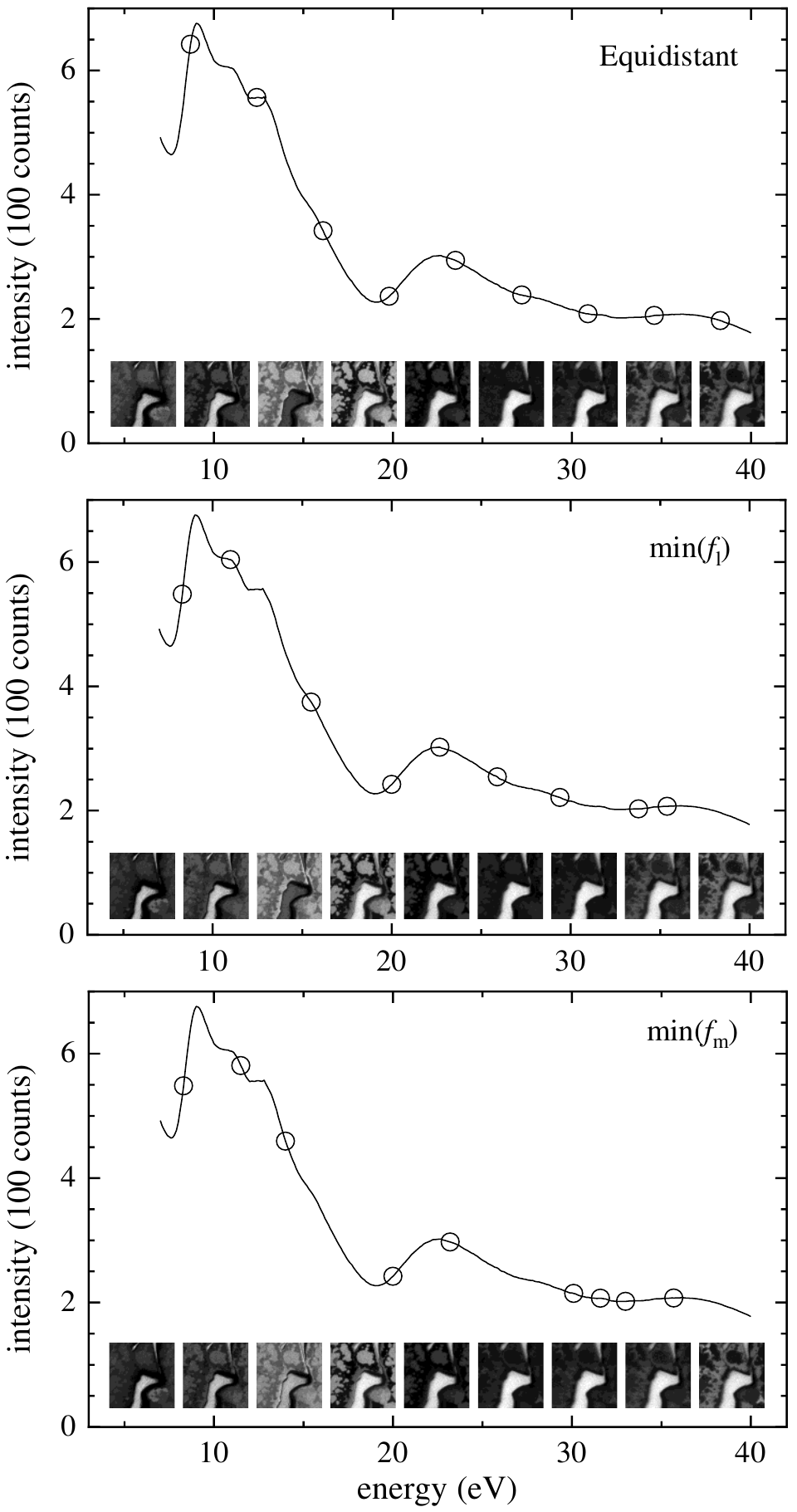}
	\caption{Spectral points (symbols) obtained by equidistant sampling (top) or minimising the FOMs, \fl\ (middle) and \fm\ (bottom). The solid line shows the spatially averaged spectrum of the data. The thumbnails show LEED intensity of a selected region at the energies of the spectral points, in the same horizontal order. $\sigtt=3$ . The results for \sigtt\ of 2 and 4 are given in the SI \Fig{S-fig:SparseSSI1} and \Fig{S-fig:SparseSSI2}, respectively.}
	\label{fig:SparseS}
\end{figure}

\subsection{Sparse sampling of \PrO}

The sparse sampling method detailed in \Sec{sec:sparse} was applied to the \PrO\ data, with results shown in \Fig{fig:SparseClass}. The FOMs are shown in \Fig{fig:SparseClass}a for different number of spectral points \Ns. We find that generally the FOMs decrease with increasing \Ns, as can be expected from the increasing information available. An approximate scaling of both FOMs as $1/\Ns$ is found. This indicates that the FOMs are scaling with the square of the noise, as the noise is dominated by shot noise scaling as $1/\sqrt{\Ns}$.

We also find that the optimization of the spectral points is reducing the FOMs stronger for smaller \Ns. This can be understood considering the increasing spectral separation of the points for equidistant sampling with decreasing \Ns, which can miss out on relevant spectral features. Adjusting the spectral positions, the points can be repositioned to sample such features better.  For \Ns\ above 100, the optimization does not affect the FOMs, while for $\Ns=9$, the minimum number required to reconstruct the 9 \FSC\ components, the optimization improves the FOMs by about a factor of two. We also notice that optimizing for one FOM, also the other improves to some extent. Examples of the random walk used to optimize the spectral points are shown in \Fig{S-fig:SparseOptSI2}-\Fig{S-fig:SparseOptSI3} for different values of \sigtt. We find that positions close to optimum are found within a few tens of iterations.

The resulting classifications for equidistant, as well as for \fm\ or \fl\ optimized positions, are shown in \Fig{fig:SparseClass} for $\Ns=9$. They can be compared with the classification using all spectral points shown in \Fig{fig:PrOxClass}. We find that most features are recovered, as also suggested by the FOM values of below 10\%. However, \class{1}, which is only present in a small number of spatial points, is mostly unclassified for the equidistant spectral points, partially classified for the \fl\ optimized spectral points, an well reproduced for the \fm\ optimization. The spectral points used are shown in \Fig{fig:SparseS} for the three cases, together with  LEEM intensity images of a selected sample region at the spectral points.

This example shows that the sparse sampling allows to reconstruct spectra and classifications with a strongly reduced number of spectral points, allowing to speed up data acquisition in the present case by a factor of 33. Notably, both the fraction of non-classified points \fl, and of misclassified points \fl, remain below 10\% even in this case, a value limited by the SNR of the data rather than genuinely missing spectral information.

\subsection{\FSC\ and classification of \RuO} \label{sec:fscclassruo}

\begin{figure}
	\centering
	\includegraphics[width=0.5\textwidth]{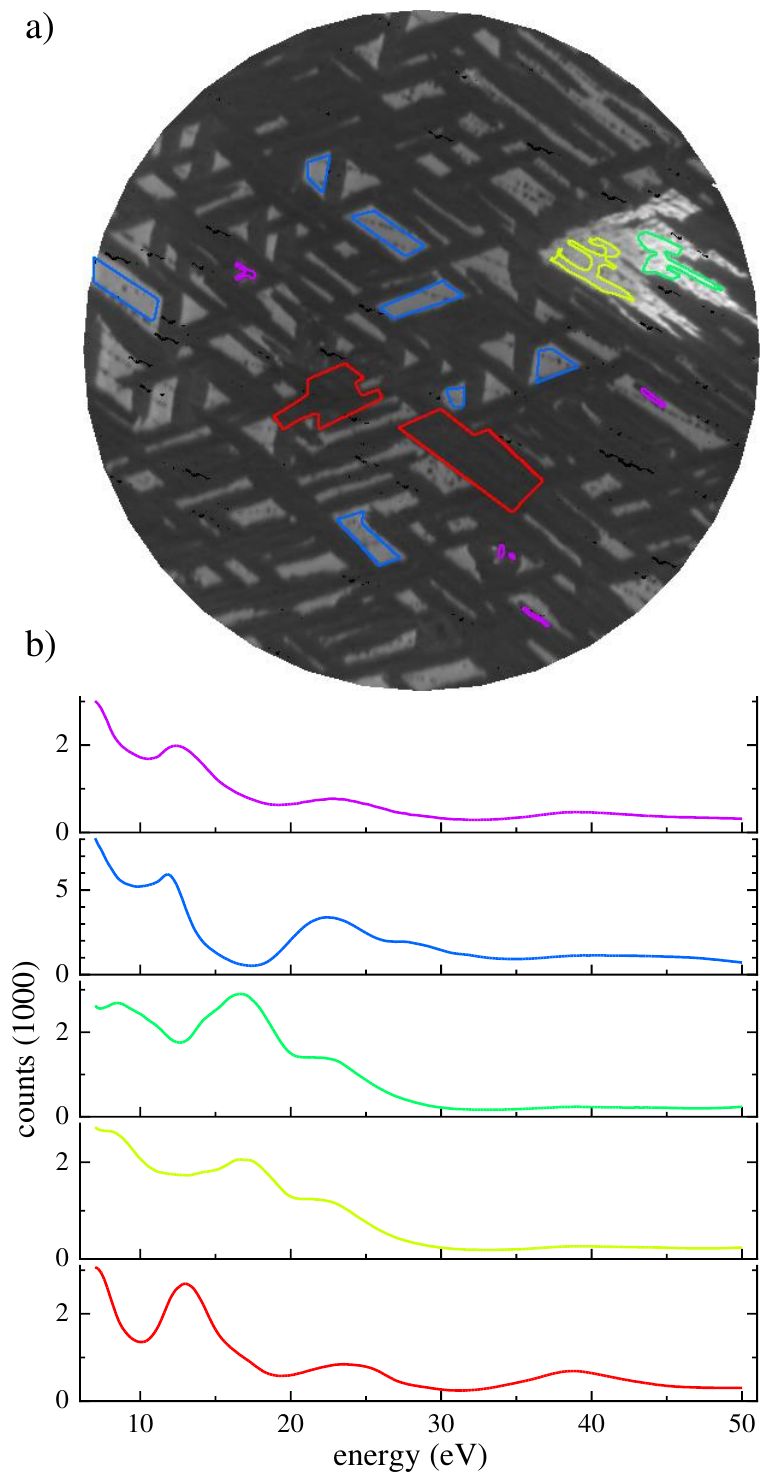}
	\caption{a) LEEM image of the oxidized \ce{Ru(0001)} surface  taken at \SI{19.0}{\eV} with a field of view of \SI{10}{\micro\meter}. b) Mean \iv\ spectra extracted from the \ce{RuO2} data by averaging over the spatial regions as indicated by encircling lines of corresponding colour in a), used as training set for the classification.
		\label{fig:RuO2_raw}}
\end{figure}

In \Fig{fig:RuO2_raw}a, a LEEM image from the \ce{Ru(0001)} surface after oxidation is presented, showing already at a first glance the different types of islands that comprise the rich \ce{RuO2}/\ce{Ru} system.
The presence of multiple different \ce{RuO2} orientations is characteristic for \ce{Ru} oxidation and has previously been observed in PEEM and LEEM \cite{bottcher2000a,flege2008}. The formation of such oxides and their application in catalysis is reviewed in \Onlinecite{over2012}. For assigning the islands and substrate regions observed to the corresponding structures, \iv\ fingerprint spectra as well as the \textmu-LEED patterns are readily available in literature.\cite{flege2015,fischer2020}

\begin{figure*}
	\centering
	\includegraphics[width=\textwidth]{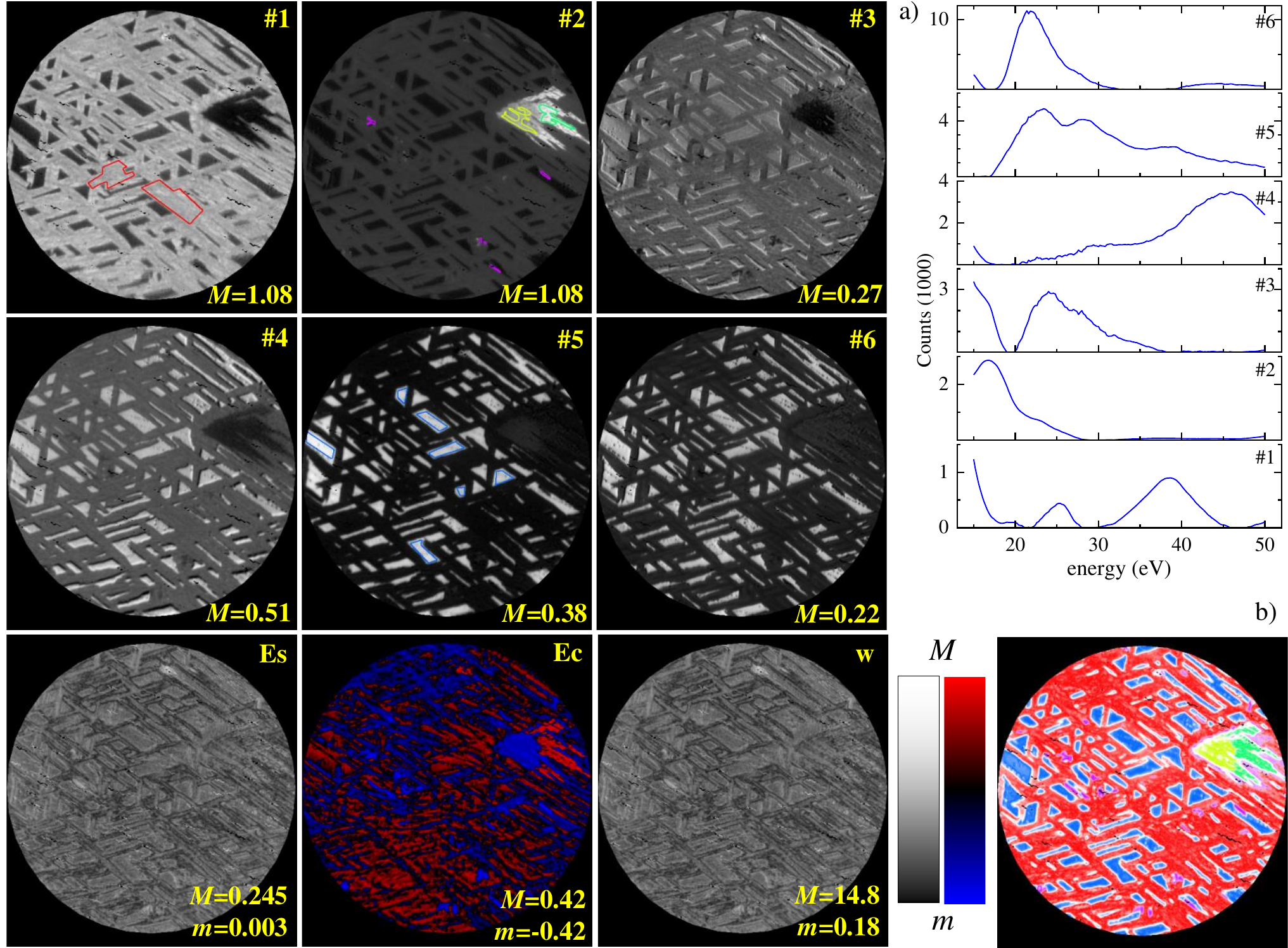}
	\caption{\FSC\ analysis and classification of LEEM \iv\ data taken on a \ce{RuO2}/\ce{Ru(0001)} surface. Concentrations of the components $\Comp_i$, $i=1..6$ are given in the images on a grayscale from $m=0$ to $M$ as labeled, together with the spectral error \Es, concentration error \Ec, and weight $w$ calculated by the weighted factorisation. \Es\ and $w$ are shown with a logarithmic scale between $m$ and $M$ as indicated. The concentration error \Ec\ is shown on a colour scale red (blue) hue for $\Ec>0$ ($\Ec<0$), with value proportional to $\log\left(\left|\Ec\right|\right)$, and black indicating $\left|\Ec\right|<0.01$. The spatial regions used as training sets for classification into five classes are indicated by lines colored according to the associated class, in the component image where they are most pronounced. a) Component spectra. b) Classification result, with the hue indicating the class, and the saturation the confidence, as in \Fig{fig:PrOxClass}. The component concentrations over the training regions and the class spectra are shown in the SI in \Fig{S-fig:RuO2Class}, with the separate $\sigt$ for each class given in \Fig{S-fig:RuO2ClassSI}.}
	\label{fig:RuO2}
\end{figure*}

\Fig{fig:RuO2} shows the \FSC\ analysis (see \Sec{sec:FSC}) of the \iv\ LEEM stack on \RuO\ using 6 components with the component spectra shown in \Fig{fig:RuO2}a. The component concentration images clearly exhibit the surface morphology expected from the single LEEM image and emphasize different parts of the surface each. As in the \ce{PrO$_x$} data, we manually select areas as indicated by the colored lines in the component concentration images and perform a classification. The result is shown in \Fig{fig:RuO2}b and the average \iv\ spectra of each training region are presented in \Fig{fig:RuO2_raw}b.

Most of the surface is covered by elongated structures that are about \SI{300}{\nano\meter} wide, \SI{1}{\micro\meter}-\SI{5}{\micro\meter} long, and are aligned along three different directions. This phase has a large $\Comp_1$ contribution. It exhibits the typical morphology of \ce{RuO2(110)} islands grown below \SI{500}{\celsius} sample temperature \cite{flege2018}: The threefold island symmetry corresponds to the formation of three different rotational domains with the \ce{RuO2}-[001] crystallographic direction aligning with the primary directions of the substrate, \ce{Ru}-$\langle 11\overline{2}0\rangle$. \cite{he2005} \textmu-LEED patterns taken in such sample regions (see \Fig{S-fig:LEED_SI} b) and the corresponding \iv\ curve (see \Fig{fig:RuO2_raw}b) confirm this. 

The classification shown in \Fig{fig:RuO2}b reliably assigns this phase in red. Only some areas at the fringe of \ce{RuO2(110)} islands show low classification confidence --  this is attributed to non-diffraction contrast mechanisms at these islands' borders due to their height---a slight beam tilt visible in $\Comp_4$ as well as some lens astigmatism recognizable in $\Comp_3$'s anisotropy of the \ce{RuO2(110)} rotational domains conspire to impair the classification accuracy.

In $\Comp_6$, some variation across the \ce{RuO2(110)} islands' width is visible with essentially two different contrast levels. $\Comp_6$ emphasizes the peak at \SI{22}{\eV}, which is characteristic for the substrate phase (see next paragraph). Variation in this component thus can be interpreted as a variation in the substrate signal, i.e., a thickness variation of the \ce{RuO2(110)}. This indicates a decreased thickness of the \ce{RuO2}(110) islands at their edges. As reported in \Onlinecite{he2005}, \ce{RuO2(110)} growth is limited to \SI{1.6}{\nano\meter} (\SI{5}{ML}) thickness below \SI{350}{\celsius} and then gradually increases with growth temperature. In our case of \SI{410}{\celsius} growth temperature, small \ce{RuO2(110)} islands at first only exhibit the contrast level associated with less thickness, arguably limited from further growth. After reaching a certain width, the island allows for vertical growth again and form a thicker core in their center region. 

On the right hand side of the data set, there is a large, arrowhead-shaped area that has a different contrast than the above mentioned ones. In the growth video, this phase nucleates last and quickly grows to fill the area between the encompassing \ce{RuO2(110)} islands. The LEED pattern (see \Fig{S-fig:LEED_SI} c) as well as the \iv\ spectrum show that this is \ce{RuO2(101)}. The island exhibits two different contrast levels as best seen in $\Comp_2$, $\Comp_3$ and $\Comp_5$. This is because the left side of the island was subjected to a much higher electron flux at higher energies (\SI{>30}{\eV}) during a \textmu-LEED measurement series. The surrounding area was not affected as the electron beam was shielded by the illumination aperture.

The ``substrate'', as mentioned, is most prominent in $\Comp_6$ and also features in several other components, most strongly in $\Comp_4$ an $\Comp_5$, where the \ce{Ru} step edge decorations are less pronounced. The signal weakest in $\Comp_1$ and $\Comp_2$, owing to those components' large emphasis on the region around \SI{17}{\eV} where the substrate \iv\ curve has a dip. In the classification image, it features as the blue phase. Based on the \iv\ curve in \Fig{fig:RuO2_raw}b and the $(1\times 1)$ reconstructed hexagonal LEED pattern (see \Fig{S-fig:LEED_SI} a), this ``substrate'' phase can be assigned unambiguously to a one-monolayer oxygen adlayer on \ce{Ru(0001)}, where each hcp hollow site is occupied by a single \ce{O} atom \cite{krasovskii2014}.

The adlayer-covered substrate areas encompass small, roundish islands (diameter \SI{\approx 10}{\nano\meter}) that decorate the original \ce{Ru} step edges. This phase has not been captured in a separate component, due to its small surface coverage. Still, its distribution on the surface is apparent in the classification image \Fig{fig:RuO2}b inside the blue ($1\times1$)-\ce{O} phase as white areas where the classification confidence is low. The close arrangement along step edges corroborates the nucleation-and-growth process at single steps laid out in \Onlinecite{herd2014} on a mesoscale. Presumably, the majority of the nuclei evolved into \ce{RuO2(100)} islands, exhibiting the characteristic roundish shape. \cite{flege2016} However, it is not possible to get conclusive LEED images from areas this small. 

Interestingly, there is still another phase as revealed by the \FSC\ analysis. Visible as highly negative values in the concentration error \Ec\ in \Fig{fig:RuO2}, there are small spots of somewhat irregular, elongated shape that exhibit a strongly reduced ``concentration''. Parts of these objects do, however, show up brightly in $\Comp_2$ (corresponding to \ce{RuO2(101)}) while their averaged \iv\ curve resembles that of \ce{RuO(110)} (see \Fig{fig:RuO2_raw}, the red curve resembles the purple one). This indicates that they contain both orientations with grains smaller than the LEEM's lateral resolution, which in first approximation leads to a linear mixing of intensities \cite{schmid1995} and thus of the corresponding \FSC\ components. The remaining concentration error then is explained by the strong faceting of some grains: Facet planes that are not parallel to the surface cannot contribute significant intensities in the specular direction, which is selected by the contrast aperture in \iv\ LEEM and represents the zeroth order of diffraction for planes normal to the optical axis. Hence, as a large surface fraction of such faceted regions remains undetected, a decrease in the overall ``concentration'' of crystalline material in the \FSC\ results for these islands. Considering these three components, the phase is identified as nanocrystalline \ce{RuO2} with diverse orientations as described by \citet{flege2015} and \citet{herd2013}.

\section{Conclusions and Outlook}

We have demonstrated a novel data analysis pipeline for LEEM \iv\ data. The data is represented by a few surface phase components and their spatial concentration maps. Details beyond this model are observable through analyzing the factorization error in both concentration and spectrum. 

For the \ce{RuO2} system, it could be shown that beside mesoscale \ce{RuO2} islands of distinct orientations, a nanocrystalline phase exists that contains \ce{RuO2(101)} and \ce{RuO2(110)} grains with lateral sizes below this LEEM instrument's resolution, i.e., \SI{<10}{\nano\meter}. 
A very convincing classification could also be achieved for the complex exemplary dataset on \PrO\, featuring a clear distinction between all pre-characterized surface phases despite of their only subtle physical differences.

Using the concentrations, we have demonstrated a supervised classification method which classifies every point on the surface into one of the training phases, or into non-classified if the mismatch to the training set is exceeding a given value of standard deviations. 

Based on this factorization, we show a sparse sampling method using the spectral components to reconstruct full spectra from a small number of spectral points followed by classification. In the example given on the \PrO\ dataset, a speed up by a factor of 33 was achieved. This opens the perspective of real time classified imaging of processes on a multi-phase surface via dynamical LEEM \iv.  Notably, the component concentrations could also be used for an unsupervised classification, which we plan to explore in a future work. Information on the data underpinning the results presented here, including how to access them, can be found in the Cardiff University data catalogue at http://doi.org/10.17035/d.2022.0153725100.

\section*{Author contributions}
W.L. and J.F. conceptualized the work and methodology. 
S.F. and J.-O.K. prepared the samples and acquired the data.
F.M. and W.L. developed the analysis software.
F.M. analyzed the data. S.F. and J.-O.K. interpreted the classification results on the exemplary datasets.
F.M., W.L., J.-O.K. and S.F. wrote the manuscript. J.F. reviewed the manuscript.

\section*{Conflicts of interest}
There are no conflicts to declare.

\section*{Acknowledgements}
F.M. acknowledges the Ser Cymru II programme (Case ID 80762-CU-148) which is funded by Cardiff University and the European Regional Development Fund through the Welsh Government.  
J.-O.K. acknowledges funding by DFG (Grant Numbers 362536548, 408002857).
J.F., J.-O.K. and W.L. acknowledge travel funding by the Bremen-Cardiff alliance.

\end{document}